
\def\({\left(}
\def\){\right)}
\def\[{\left[}
\def\]{\right]}
\def\a{\alpha}

\def\z{\zeta}

\def\coma{\quad ,\quad}
\def\frac#1#2{{#1\over#2}}

\def\vspace#1{\crcr\noalign{\vskip#1\relax}}
\def\sqr#1#2{{\vcenter{\vbox{\hrule height.#2pt
                        \hbox{\vrule width.#2pt height#1pt \kern#1pt
            \vrule width.#2pt}
         \hrule height.#2pt}}}}

\def\wave#1{\mathaccent "7E #1}
\catcode`\@=11
\def\displaylinesno #1{\displ@y\halign{
\hbox to\displaywidth{$\@lign\hfil\displaystyle##\hfil$}&
\llap{$##$}\crcr#1\crcr}}
\catcode`\@=12

\magnification= \magstep1
\hsize=4.9in
\vsize=7.35in
\hoffset=14truemm
\baselineskip 14 pt
\overfullrule 1pt

\centerline{\bf  APPROACHES TO THE MONOPOLE--DYNAMIC
DIPOLE }
\vskip 2mm
\centerline{\bf VACUUM SOLUTION CONCERNING THE STRUCTURE OF  }
\vskip 2mm
\centerline{\bf  ITS
ERNST'S POTENTIAL ON THE SYMMETRY AXIS}
\bigskip
\medskip
\centerline{J. L. Hern\'andez-Pastora$^\dagger$, J. Mart\'\i
n$^\dagger$, E. Ruiz \footnote{$^\dagger$}{Area de F\'\i sica Te\'orica.
 Edificio Triling\" ue, Universidad de Salamanca. 37008 Salamanca, Spain} }

\medskip

\vskip  1cm

{\bf Abstract} \quad The FHP algorithm [1] allows to obtain the
relativistic multipole moments of a vacuum stationary 
axisymmetric solution in terms of coefficients which
appear in the expansion of its Ernst's potential $\xi$ on the
symmetry axis. First of all, we will use this result in order to
determine, at a certain approximation degree, the Ernst's potential on
the symmetry axis of the metric whose only multipole moments
are mass and angular momentum. 

By using Sibgatullin's method [2] we then analyse a
series of exact solutions with the afore mentioned multipole
characteristic; besides, we present an approximate solution
whose Ernst's potential is introduced as a power series of a
dimensionless parameter. The calculation of its multipole moments allows us to
understand  the existing differences between both
approximations to the proposed pure multipole solution.
\vskip  1cm

\centerline{Submitted to {\it General Relativity and Gravitation}}

\vskip  1cm
Jos\'e Luis Hern\'andez-Pastora

 Max-Planck-Institut f\" ur Gravitationsphysik.

Albert-Einstein-Institut.

 Schlaatzweg 1, D-14473 Potsdam (Germany)

Phone: +49 - 331 - 2753748

Fax: +49 - 331 - 2753798

e-mail: pastora@aei-potsdam.mpg.de

\vfill\eject

\noindent {\bf 1. INTRODUCTION}

\vskip 0.5cm

As it is well known, the relativistic multipole moments defined by
Geroch [3] and Hansen [4] allow us to  characterize, in an
unique manner, vacuum stationary metrics. Particularly, 
Schwarzschild's solution (spherically symmetric) can
be described as that one whose unique multipole
moment is monopole.

In [5] we have introduced a static and axisymmetric solution
of the Einstein vacuum field equations whose only multipole
moments are mass and the quadrupole moment. Our aim in the present
work is to generalize the above mentioned result to the non-static case,
by searching the stationary and axisymmetric vacuum
solution  which in addition to mass possesses only a dynamic
dipole moment.

Fodor, Hoenselaers and Perjes [1] have developed  an algorithm to
calculate multipole moments in terms of the  coefficients arising
in the expansion of the Ernst's potential  $\xi$  on the symmetry axis
as a power series on the inverse of the Weyl's coordinate $z$. According
to this result it is possible to determine such coefficients
corresponding to a solution with the mentioned multipole
characteristics.

There is a method, due to Sibgatullin [2], which allows to
generate exact solutions of vacuum, stationary and axisymmetric
field equations from the Ernst's potential on the symmetry axis,
whenever its structure is a polynomial ratio. This method
has been broadly  used [6], [7], [8] and
 it has been recently completed by the introduction of some general
expressions which make its use much easier [9].

Our first aim, is to prove that a finite set of
coefficients which appear in the expansion of the potential $\xi$ on
the symmetry axis is sufficient enough to describe a potential $E
\equiv(1-\xi)/(1+\xi)$ of rational type on the symmetry axis.
Therefore, the use of  Sibgatullin's method, makes it possible to
construct an Ernst's potential $E$ in terms of these
coefficients. Nevertheless, if we look for an exact solution of the
$M-J$ type, then the result of this condition leads to the Ernst's potential 
on the symmetry axis by means of a ratio of series.
Hence, there is no finite number of coefficients which
describe the Ernst's potential of Monopole-Dynamic dipole solution
as a polynomial ratio. 

We then approach  the $M-J$ solution  as the limit of a sequence
of exact solutions which possess a progressively smaller number of
multipole moments higher than the dipole. In spite of this,
the behaviour of the multipole moments shows that the mentioned series
approaches the $M-J$ solution in a rather unexpected way.

In order to obtain an alternative approach to the $M-J$ solution,
we will propose a series of approximate solutions described by the
partial sums of the expansions of the Ernst's potential $\xi$ as a power
series in a dimensionless  parameter
${\cal J}$. To do that, we use  Schwarzschild's solution as a seed
solution and solve Ernst's equation at successive orders in the parameter and
impose the corresponding structure of the $M-J$ solution on the symmetry
axis.

As we will see, the order of magnitude of its multipole moments
decreases as the order of approximation rises. That leads us to conclude
that the parameter
${\cal J}$ controls the successive
solutions and contributes to give a physical meaning to the approximate
solution. 

A more realistic stationary solution would be a $M-Q-J$ solution,
i.e., the one having only mass, massive quadrupole moment and
dynamic dipole, since a rotating object flattens and,
hence, all its massive multipole moments represent such deviation
from sphericity. Nevertheless, we can imagine so rigid  an object
 that the $M-J$ solution  itself would be physically relevant.
Besides, since the static case is already solved by $M-Q$ solution, we
want to discuss $M-J$ solution  and consider the $M-Q-J$ solution 
 as a  generalization of both solutions.

\vskip 1cm

\noindent {\bf 2. STRUCTURE OF THE M--J SOLUTION ON THE
SYMMETRY AXIS}

\vskip 0.5cm

Let us be $\xi$ the Ernst's potential of a stationary 
axisymmetric solution of vacuum Einstein's field equations [10]
$$
(\xi \xi^*-1) \triangle \xi = 2 \xi^* (\nabla \xi)^2 \coma
\eqno(1)$$
being $\displaystyle \xi \equiv \frac {1-E}{1+E}$, where the
Ernst's potential $E$ is the complex function whose real part
represents the norm of the Killing vector describing
stationarity. On symmetry axis, this potential $\xi$ can be expanded
by means of a power series of the inverse Weyl's coordinate $z$ as follows:
$$
\xi (\rho=0,z) = \sum_{n=0}^{\infty} m_n z^{-(n+1)}
\eqno(2)
$$
where $\rho$ represents the Weyl's radial coordinate.

Fodor, Hoenselaers and Perjes [1] have developed an algorithm which
allows to calculate the Geroch [3] and Hansen [4] relativistic
multipole moments, related to a vacuum stationary 
axisymmetric solution, in terms of the coefficients $m_n$ arising
in the previous expansion (2). Both the result obtained up to
multipole order $10$ by the afore mentioned authors, and  the
calculations we have carried out up to order $20$ lead us to show that
the relation between multipole moments and  coefficients $m_n$ is
triangular. That is to say, the multipole moment and the corresponding
coefficient $m_n$ at every order, differ in a certain combination of
lesser order  $m_k$ coefficients. Therefore, these relations enable
us to determine unequivocally the coefficients $m_h$  which allows to outline
 the expansion of $\xi$ in terms of the known multipole
moments for any given solution.

So  we have obtained that the solution having only massive
monopole and dynamic dipole is characterized by an Ernst's potential
$\xi$ whose expansion on the symmetry axis provides the following 
coefficients $m_n$ up to the order $20$,
$$
\eqalign{
m_0 &= M \coma m_1 = i J \cr
\vspace{2mm}
m_2 &= 0 \coma m_3 = 0 \cr
\vspace{2mm}
m_4 &= \frac1 7 M J^2 \cr
\vspace{2mm} 
m_5 &=  -\frac{1}{21} i J^3 \cr
\vspace{2mm} 
m_6 &=  \frac{1}{21} M^3 J^2 \cr
\vspace{2mm} 
m_7 &=  -\frac{13}{231} M^2 i J^3 \cr
\vspace{2mm}
m_8 &=  \frac{5}{231} M^5 J^2-\frac{40}{3003} M J^4 \cr
\vspace{2mm} 
m_9 &= -\frac{115}{3003} M^4 J^3 i-\frac{4}{3003} J^5 i \cr
\vspace{2mm} 
m_{10} &=  \frac{5}{429} M^7 J^2-\frac{115}{7007} M^3 J^4 \cr
\vspace{2mm}
m_{11} &=  -\frac{1}{39} M^6 J^3
i-\frac{389}{357357} M^2 J^5 i \cr
\vspace{2mm} 
m_{12} &= \frac{1}{143} M^9 J^2-\frac{1569}{119119} M^5
J^4-\frac{53}{2909907} M J^6 \cr 
\vspace{2mm}
m_{13} &= -\frac{43}{2431} M^8 J^3
i-\frac{265}{108927} M^4 J^5 i-\frac{13051}{20369349} J^7 i \cr
\vspace{2mm} 
m_{14} &= \frac{1}{221} M^{11} J^2-\frac{187618}{20369349} M^7 J^4+
\frac{1129}{10968111} M^3 J^6 \cr
\vspace{2mm} 
m_{15} &= -\frac{53}{4199} M^{10} J^3
i-\frac{10954}{2263261} M^6 J^5 i-
\frac{831513}{364385021} M^2 J^7 i \cr
  }
$$

\vfill\eject

$$
\eqalign{
m_{16} &=  \frac{1}{323} M^{13}
J^2-\frac{40346}{6789783} M^9 J^4-
\frac{454}{15954939} M^5 J^6-\frac{2419504}{16397325945} M J^8 \cr
m_{17} &=  -\frac{3}{323} M^{12} J^3
i-\frac{12480070}{1717815099} M^8 J^5 i-
\frac{293822614}{60123528465} M^4 J^7 i- \cr
\vspace{2mm} 
& -\frac{35634548}{147575933505} J^9 i \cr
\vspace{2mm} 
m_{18} &=  \frac{5}{2261}
M^{15} J^2-\frac{862123}{245402157} M^{11} J^4-
\frac{10703470}{36074117079} M^7 J^6- \cr
\vspace{2mm} 
& -\frac{11186023022}{21103358491215} M^3 J^8 \cr
\vspace{2mm}
m_{19} &=  -\frac{173594465986}{21103358491215} M^6 J^7 i-
\frac{99041865574}{87428199463605} J^9 M^2 i-
\frac{365}{52003} M^{14} J^3 i - \cr
\vspace{2mm} 
&-\frac{6808829}{736206471} M^{10}
J^5 i \cr
\vspace{2mm} 
m_{20} &=  \frac{5}{3059} M^{17} J^2-\frac{6540151}{3681032355}
M^{13} J^4-\frac{9908406983}{21103358491215} M^9 J^6- \cr
\vspace{2mm} 
& -\frac{4674899812546}{4283981773716645} M^5 J^8-
\frac{985078066594}{18971919283602285} M J^{10} \coma\cr } 
\eqno(3)
$$
where $M$ and $J$ represent Mass and Angular Momentum respectively.

These expressions suggest that the coefficients $m_n$ of the potential 
$\xi$ representing the Monopole-Dynamic dipole solution  can be
written in the following way
$$
\eqalign{
&m_{2k} = M^{2k+1} \sum_{n=1} ^{(2k+1)/4} {\cal J}^{2n}\,
G(2n,2k) \cr
\vspace{2mm} 
&m_{2k+1} = M^{2k+2} \sum_{n=1}^{k/2} {\cal J}^{2n+1}\,
G(2n+1,2k+1)\cr
} \coma (k=0,1,\dots)
\eqno(4)$$
where we have introduced a dimensionless parameter
$\displaystyle {\cal J} \equiv \frac{m_1}{m_0^2} = i \frac{J}{M^2}$
and the function $G(l,h)$. For every coefficient $m_h$ this function 
describes the numerical factor multiplying the  power $l$ of the parameter
${\cal J}$.

By substituting the  expressions 
(4) in the expansion (2) of the Ernst's potential $\xi$ on the symmetry
axis, and rearranging sums, it is possible to write this potential
as a power series of the parameter ${\cal J}$:
$$
\xi(\rho=0,z)=  \frac{M}{z} +  {\cal J}\frac{M^2}{z^2} + \sum_{\a
=2}^{\infty} {\cal J}^{\a}
\Phi_{\a} \qquad ,
\eqno(5)$$
where functions $\Phi_{\a}$ are defined below
$$
\eqalignno{
\Phi_{2n} = & \sum_{k=2n}^{\infty} G(2n,2k) \hat \lambda^{2k+1} &
(6a) \cr
\vspace{2mm}
\Phi_{2n+1} = & \sum_{k=2n}^{\infty} G(2n+1,2k+1) \hat
\lambda^{2k+2} \quad , & (6b) \cr
}
$$
with the notation $\displaystyle \hat \lambda \equiv \frac Mz$.
Let us note that since the parameter ${\cal J}$ is imaginary,
the functions $\Phi_{\a}$ with an odd index turns the series (5) into 
an imaginary function.

As shown in (5), the Ernst's potential $\xi$ can be expressed on the
symmetry axis by a double series; one of them in terms of
the parameter ${\cal J}$ and the other being a power series of the inverse
coordinate $z$. Nevertheless, as we will see, 
 the sum of the series
$\Phi_{\a}$ can be obtained, at least up to  first orders. In order to
do that it is necessary to obtain the analytic expressions which
describe the double index functions 
$G(\a,h)$. If the first of those indexes is fixed, that is to say, if we 
consider a certain value for the power of the parameter
${\cal J}$, we have tried to adjust the resulting series of
the corresponding terms arising from every coefficient $m_n$.

For example, it is very easy to check that factors appearing with
powers two and three in the parameter ${\cal J}$ verify
respectively the following expressions 
$$
\eqalign{
&G(2,2k) = \frac{15}{(2k+3)(2k+1)(2k-1)} \cr
\vspace{2mm} 
&G(3,2k+1) = \frac{15(10k-17)}{(2k+5)(2k+3)(2k+1)(2k-1)}\cr
}
\eqno(7)$$
Now, it is quite simple to obtain the sum of the series $\Phi_{\a}$
by rewriting the functions $G(\a,h)$ as a sum of irreducible fractions.
Particularly, for $G(2,h)$ we have
$$
\eqalign{
& G(2,2j+1) = 0 \cr
\vspace{2mm}
& G(2,2j) =  \frac{15}{8}
\[\frac{1}{2j+3}-\frac{2}{2j+1}+\frac{1}{2j-1}\] \equiv
\sum_{i=0}^2\frac{g_i^{(2)}}{2i+2j-1} \cr
}
\eqno(8)$$
Taking this expression into (6a) the function $\Phi_2$ gives
$$
\Phi_{2} =  \sum_{k=2}^{\infty}\,
\sum_{j=0}^2 \frac{g_j^{(2)}}{2k+2j+1} \hat \lambda^{2k+1} \quad .
\eqno(9)
$$
Rearranging sums and making use of Lemma 3 of Appendix,
we obtain the following finite sum:
$$
\Phi_{2} = \sum_{j=0}^2 g_j^{(2)} \hat \lambda^{4}
\sum_{k=0}^{3-j} C_{2(3-j),2k} Q_{2k}(1/ \hat \lambda) \coma
\eqno(10)$$
where the coefficients $C_{lh}$ are defined in Appendix and the functions 
$Q_h(x)$ are special Legendre's functions of second kind. Let us note
that the previous expression can be written as follows
$$
\Phi_{2} = \hat
\lambda^{4} \sum_{k=0}^{3} Q_{2k}(1/\hat \lambda)
\sum_{j=0}^{min(3-k,2)} g_j^{(2)} C_{2(3-j),2k} \quad .
\eqno(11)$$

\vskip 1cm

\noindent {\bf 3. SEQUENCE OF EXACT SOLUTIONS}

\vskip 0.5cm

The expression of Ernst's potential on the symmetry axis can be
used as boundary condition to obtain solutions of the Ernst's equation.
For example, Sibgatullin's method [2] simplifies this problem
by solving a linear system of integral equations. So, the Ernst's
potential $E$ results from the following expression
$$
E= \frac 1{\pi} \int_{-1}^1 \frac{\mu(\sigma)
e(\tau)}{\sqrt{1-\sigma^2}} d\sigma \coma
\eqno(12)
$$
where $\tau$ is a complex variable defined from the cilindrical Weyl's
coordinates  $\tau\equiv z+i\rho\sigma$, and $\sigma \in [-1,1]$
is an arbitrary integration variable. Function $e(z)$
represents the value of the Ernst's potential $E$ on the symmetry
axis, that means, $e(z) \equiv E(\rho=0,z)$. At length, the function
$\mu(\sigma)$ must be a solution verifying the following integral
equations system
$$
\eqalignno{
& \wp\int_{-1}^1
\frac{h(\tau,\eta)\mu(\sigma)}{(\tau-\eta)\sqrt{1-\sigma^2}}
d\sigma = 0 & (13a)\cr
\vspace{2mm}
& \frac 1{\pi} \int_{-1}^1 \frac{\mu(\sigma)}{\sqrt{1-\sigma^2}}
d\sigma = 1 \qquad, & (13b) \cr}
$$
being $\eta$ a complex variable defined as $\eta
\equiv z+i \rho \varsigma$, with $\varsigma \in [-1,1]$, and where
the symbol $\wp$ stands for the taking of the principal part of the
integral. On the other hand, the function $h(\tau,\eta)$ is defined as 
is shown below
$$
h(\tau,\eta) \equiv e(\tau)+\wave{e}(\eta) \coma
\eqno(14)
$$
where the function $\wave{e}(\eta)$ is obtained from $e(\eta)$
by conjugating first the variable, $\eta\rightarrow\eta^*$, and then
 the function, i.e.,
$
\wave{e}(\eta) = e^*(\eta^*) $.

Obviously, the general solution of the equations in (13) is not
evident. Nevertheless, rather compact expressions have been
obtained [9] for the Ernst's potential when the boundary
condition $e(z)$ is a rational function, i.e, 
$$
E(\rho=0,z) \equiv e(z) = \frac{P(z)}{Q(z)} \coma
\eqno(15)
$$
where $P(z)$ and $Q(z)$ are polynomials in the variable $z$, which,
taking into account that the Ernst's potential must tend to $1$ in the
neighbourhood of infinity, should be as follows
$$
\eqalign{
P(z) = & z^N+\sum_{k=1}^N a_k z^{N-k} \cr
Q(z) = &  z^N+\sum_{k=1}^N b_k z^{N-k} \cr}\quad .
\eqno(16)
$$

Since we know the structure of the potential $\xi$ on the symmetry
axis in terms of coefficients $m_n$, inmediately a
question arises: wether there exists a  relation between these
coefficients and those of the polynomials in (16). The above question 
has been answered in [11] and in what follows we give the resulting
expressions for $P(z)$ and $Q(z)$ in terms of $m_n$:
$$
P(z) = (L_N)^{-1} \left |
\matrix{\vspace{4mm}
z^N-\sum_{n=0}^{N-1}m_n z^{N-1-n} & m_N &
\dots & m_{2N-1} \cr
\vspace{4mm}
z^{N-1}-\sum_{n=0}^{N-2} m_n z^{N-2-n} & m_{N-1}  & \dots &
m_{2N-2}
\cr
\vspace{4mm}
\dots & \dots & \dots  & \dots  \cr
z-m_0 & m_1 &  \dots &  m_{N} \cr
1 & m_0 & \dots  & m_{N-1}  \cr} \right | \coma
\eqno(17)
$$
$$
Q(z) = (L_N)^{-1} \left |
\matrix{\vspace{4mm}
z^N+\sum_{n=0}^{N-1}m_n z^{N-1-n} & m_N &
\dots & m_{2N-1} \cr
\vspace{4mm}
z^{N-1}+\sum_{n=0}^{N-2} m_n z^{N-2-n} & m_{N-1}  & \dots &
m_{2N-2}
\cr
\vspace{4mm}
\dots & \dots & \dots  & \dots  \cr
z+m_0 & m_1 &  \dots &  m_{N} \cr
1 & m_0 & \dots  & m_{N-1}  \cr} \right | \coma
\eqno(18)
$$
where the following determinant $L_N$ has been defined:
$$
L_N \equiv \left | \matrix{m_{N-1} & m_{N} & \dots & m_{2N-2} \cr
  m_{N-2}   &       m_{N-1} & \dots & m_{2N-3} \cr
      \dots & \dots & \dots & \dots \cr
m_{0} & m_{1} & \dots & m_{N-1} \cr} \right |.
\eqno(19)
$$

Hence, from $2N$ coefficients $m_k$, it is possible to build on the symmetry
 axis the Ernst's potential $E$
of a vacuum solution, which is a ratio of order $N$
polynomials.

Let us consider now the following question: is  potential $E$
of the Monopole-Dynamic dipole solution a polynomial
ratio on the symmetry axis?.  On handling the coefficients $m_n$ in (3)
corresponding to such solution, 
the determinants $L_n$ (19) seem to be unlikely to be
zero originating
from some order $N$ onwards. Since the coefficients $m_n$ are 
not available for every order,
we can only assert that the behaviour of the Ernst's 
potential of such solution  does not correspond to a ratio of
polynomials of order $N \leq 10$.

In spite of last statement, it is possible to construct a
set of rational type potentials $E$ on the symmetry
axis involving these coefficients $m_n$. Thus, by
Sibgatullin's method one can obtain  a sequence of exact solutions 
which, as will be shown, approach the 
Monopole-Dynamic dipole solution.

The coefficients $m_n$ (3) have been obtained on condition that
the multipole moments higher than the dipole are zero. Hence, on
fixing $N$ coefficients $m_n$, we will get an Ernst
potential which describes a solution  whose $N-2$
multipole moments higher than Angular momentum are
zero. At the same time, multipole moments of higher order,
although different from zero, are determined by just those $N$
coefficients $m_k$.

In order to perform the sequence of exact solutions, let us 
proceed to consider the Ernst's potential on the symmetry axis as a
ratio of polynomials whose order $N$ will be
progressively increased. Hence, at each stage, we will be fixing an 
increasingly 
bigger $2N$ number  of multipole moments for the
corresponding solution.

\vskip 0.5 cm

\noindent {\bf A) ORDER \ $N=1$}

\vskip 0.2 cm

Let us suposse first that Ernst's potential  on the
symmetry axis is  a ratio of polynomials of order
$N=1$, i.e.,
$$
e^{(1)}(z) = \frac{z+a_1}{z+b_1} \equiv
\frac{P^{(1)}(z)}{Q^{(1)}(z)} \quad .
\eqno(20)
$$
In order to calculate coefficients $a_1$ and $b_1$ we handle the
two first coefficients $m_k$ in (3),
$$
m_0 \equiv M \coma m_1 \equiv i J \quad .
\eqno(21)
$$
We obtain polynomials $P^{(1)}(z)$ and $Q^{(1)}(z)$ in terms
of these coefficients by using expressions (17) and (18)
respectively, and as a result the Ernst's potential is written on the
symmetry axis in the following way 
$$
e^{(1)}(z) = \frac {z-M-i J/M}{z+M-i J/M} \quad .
\eqno(22)
$$
The previous expression is exactly the corresponding potential of Kerr's 
metric with parameters $M$ and $a\equiv J/M$. By using
Sibgatullin's method and the expressions in [11], it is possible to 
obtain the Ernst's potential
for every range of Weyl's coordinates $\{\rho, z\}$, as follows 
$$
E^{Kerr} = \frac{ \a (r_++r_-)+i
a(r_+-r_-)-2 \a M}{\a
(r_++r_-)+i a(r_+-r_-)+2 \a M} \quad ,
\eqno(23)
$$
being $\a$ the positive root of polynomial
$P(z)\wave{Q}(z)+\wave{P}(z)Q(z)$ (numerator of the function
$h(z,z)$ (14)), i.e.,
$$
\a_{\pm} = \pm \sqrt{M^2-a^2} 
\eqno(24)
$$
and where $r_{\pm}=\sqrt{{\rho}^2+(z-\a_{\pm})^2} 
$

Now, let us calculate the coefficients $m_k$ higher than those proposed in
 (21). To do that, we can use some expressions in [11]
which relate such coefficients to those of the
polynomials $P^{(1)}(z)$ and $Q^{(1)}(z)$, and gives
$$
m_k = M  (i a)^k \equiv M^{k+1} {\cal J}^k \quad .
\eqno(25)
$$

It must be remembered  that one property of Kerr's metric turns out to
be  the identity between its multipole moments and
the coefficients $m_k$ entering the expansion of the potential $\xi$
on the symmetry axis [1], and leads to,
$$
\matrix{
M_0 =   M \coma &
M_1 =  {\cal J} M^2  \coma &
M_2 =  {\cal J}^2 M^3  \coma &
M_3 =  {\cal J}^3 M^4  \coma \cr
\vspace{2mm}
M_4 =  {\cal J}^4 M^5  \coma &
M_5 =  {\cal J}^5 M^6  \coma &
M_6 =  {\cal J}^6 M^7 \quad .\cr}
\eqno(26)
$$
Obviously, the coefficients $m_n$ higher than $m_2$ do not equal 
the corresponding coefficients of the $M-J$ solution. That is a
good reason to step forward.

\vskip 0.5cm

\noindent {\bf B) ORDER \ $N=2$}

\vskip 0.2cm

Let us consider the Ernst's potential on the symmetry axis as
a ratio of polinonials of order $N=2$, i.e.,
$$
e^{(2)}(z) = \frac{z^2+a_1 z+a_2}{z^2+b_1 z+b_2} \equiv
\frac{N^{(2)}(z)}{D^{(2)}(z)} \quad .
\eqno(27)
$$
We introduce four coefficients $m_k$, according to expressions
(3) with the following values
$$
m_0 \equiv M \coma m_1 \equiv iJ \coma m_2=m_3=0 \quad .
\eqno(28)
$$
If we choose the coefficients $m_k$ in (28), the solution we
 generate will have the quadrupole moment and the octupole moment 
equal to zero since $m_2$ and
$m_3$ are just equal to these multipole moments
respectively, .

The calculations of the polynomials $P^{(2)}(z)$ and $Q^{(2)}(z)$ leads to
the following result
$$
e^{(2)}(z) = \frac{z^2-Mz-iJ}{z^2+Mz+iJ} \quad .
\eqno(29)
$$

In order to construct the Ernst's potential  according to 
Sibgatullin's method  it
is necessary to obtain the roots
of the function $h(z,z)\equiv P(z)\wave{Q}(z)+\wave{P}(z)Q(z)$.
For this case, two roots of that function turns out to be real
numbers  whereas the other two are imaginary  conjugated numbers:
$$
\eqalign{
\a_1^{\pm} = & \pm \frac M{\sqrt 2} \sqrt{1+\sqrt{1-4{\cal
J}^2}} = \pm \frac M2 \[ \sqrt{1+2 {\cal J}}+\sqrt{1-2
{\cal J}} \] \cr
\a_2^{\pm} = & \pm \frac M{\sqrt 2} \sqrt{1-\sqrt{1-4{\cal
J}^2}} = \pm \frac M2 \[ \sqrt{1+2 {\cal J}}-\sqrt{1-2
{\cal J}} \] \cr}\quad .
\eqno(30)
$$
From these roots it is possible [11] to write out the Ernst's potential
$E$ as follows
$$
E^{(2)}(\rho,z) \equiv \frac{\Lambda+\Gamma}{\Lambda-\Gamma} \coma
\eqno(31)
$$
\vskip 3mm
$$
\eqalign{
\Lambda \equiv  \ &  \frac 18 r_1^+ \[ \(\frac{p-1}{p}\)^2
r_2^++\(\frac{\bar p+1}{\bar p}\)^2 r_2^-\]\cr
\vspace{2mm}
&+\frac 18 r_1^- \[\(\frac{\bar p-1}{\bar p}\)^2
r_2^++\(\frac{ p+1}{p}\)^2 r_2^-\] +\frac 14 \frac{p^2\bar p^2-1}{p^2
\bar p^2} (r_1^-r_1^++r_2^-r_2^+) \cr
\vspace{5mm}
\Gamma  \equiv \  & \frac M8 (p-\bar p) \[
\(\frac{p-1}{p}\)\(\frac{\bar p+1}{\bar p}\)
r_1^+-\(\frac{p+1}{p}\)\(\frac{\bar p-1}{\bar p}\) r_1^- \]\cr
\vspace{2mm}
& +\frac M8 (p+\bar p) \[ \(\frac{p-1}{p}\)\(\frac{\bar
p-1}{\bar p}\) r_2^+-\(\frac{p+1}{p}\)\(\frac{\bar
p+1}{\bar p}\) r_2^- \]\coma\cr
} 
\eqno(32)
$$
where the following notation have been used
$$
\eqalign{
& p \equiv +\sqrt{1+2 {\cal J}} \cr
\vspace {3mm}
& r_i^{\pm} \equiv +\sqrt{\rho^2+(z-\a_i^{\pm})^2} \cr}
\quad .
\eqno(33)
$$

Oddly enough, for this case the structure of the 
potential $\xi$ on symmetry axis is defined by the following
coefficients $m_n$
$$
m_0 \equiv M \coma m_1 \equiv i J \coma m_k=0 \coma
\forall k\geq 2 \quad .
\eqno(34)
$$
and so, reproduce only up to order $4$ the coefficients that
characterize the $M-J$ solution. Multipole moments of this
solution turn out to be
$$
\eqalign{
M_0 & = M \cr
M_1 & = {\cal J} M^2 \cr
M_2 & = 0 \cr
M_3 & = 0 \cr     
M_4 & =  \frac 17 {\cal J}^2 M^5 \cr
M_5 & = - \frac 3{21} {\cal J}^3 M^6 \cr
}\qquad\qquad
\eqalign{
M_6 & = -\frac 1{33} {\cal J}^2 M^7 \cr
M_7 & =   \frac {19}{429} {\cal J}^3 M^8 \cr
M_8 & =  M^9 \(\frac 1{143} {\cal J}^2
-\frac{53}{3003} {\cal J}^4 \) \cr
M_9 & =  M^{10} \(- \frac{43}{2431}
{\cal J}^3 + \frac{41}{17017} {\cal J}^5 \) \cr
M_{10} & =   M^{11} \(-\frac 7{4199} {\cal
J}^2+\frac{202}{12597} {\cal J}^4 \) \cr}
\eqno(35)
$$

It should be noticed that quadrupole moment and octupole
moment are zero by contruction. The first multipole moment
different from zero is $M_4$, which turns out to be proportional to
${\cal J}^2$, one order higher than the angular momentum in the
parameter ${\cal J}$ . It is noteworthy that all higher
massive moments are proportional to ${\cal J}^2$  and the dynamic 
moments turn out to be of order $3$ in that parameter, and  therefore,
 just the same order than quadrupole
moment and octupole moment in the previous case $N=1$. We will
discuss this issue later.

Another interesting property of this solution is its equatorial
symmetry which can be inferred from the fact that its odd
multipole moments are imaginary quantities (and so, according to
notation $FHP$ [1], represent dynamic moments ) while even
multipole moments are real quantities (masive moments). It is an
intrinsec characteristic of the procedure used to construct solutions
from rational type Ernst's potentials on the symmetry axis. In fact,
it can be proved  that if coefficients $m_n$ introduced are
alternatively real and imaginary quantities, then 
the resulting Ernst's potential has equatorial symmetry. That
occurs because the coefficients $a_k$ and $b_k$ (16) fulfill the next
relation $a_k=(-1)^k \ b^*_k$. Hence, in order to have equatorial
symmetry an axisymmetric stationary and
asymptoticaly flat  vacuum solution must meet the following necessary 
sufficient condition [12], [13]:
$$
e_+(z) \ e^*_+(-z) \ =\ 1 \quad ,
\eqno(36)
$$
 where $e_+(z)$ denotes the Ernst's potential on the positive region of
the symmetry axis and symbol $ ^*$ denotes complex conjugation.

\vskip 0.5cm

\noindent {\bf C) ORDER \ $N=3$}

\vskip 0.2cm

In this case we introduce six coefficients $m_k$ according to the
ones possesing the $M-J$ solution (3), which means, 
$$
\eqalign{
& m_0 \equiv M \coma m_1 \equiv iJ \coma m_2=m_3=0 \coma \cr
& m_4 =\frac 17 M J^2 \coma m_5 = -i \frac 1{21} J^3 \ .\cr}
\eqno(37)
$$
This choice ensures that the multipole moments lesser 
than $M_6$ are zero, except for mass and angular
momentum. 

Then, the Ernst's potential on the symmetry axis can be written as
follows
$$
e^{(3)}(z) =\frac{P^{(3)}(z)}{Q^{(3)}(z)} \quad ,
\eqno(38)
$$
being 
$$
\eqalign{
 P^{(3)}(z) = & z^3+M z^2
(-1+{\cal J}/3)+z M^2 (-1/7-4{\cal J}/3)+\cr
& + M^3 (1/7+{\cal J}/7-{\cal J}^2/3) \cr
Q^{(3)}(z) = & z^3+M z^2 (1+
{\cal J}/3)+z M^2 (-1/7+4{\cal J}/3)+\cr
& + M^3 (-1/7-{\cal
J}/7-{\cal J}^2/3) \quad .\cr}
\eqno(39)
$$
The roots of the function  $h(z,z)$ (14) corresponding to this potential
 are
$$
\eqalign{
\a_1^{\pm} = & \pm M \sqrt{A^-+A^++\frac 37+\frac{{\cal J}^2}{27}}
\cr
\a_2^{\pm} = & \pm M \sqrt{-\frac 12(A^-+A^+)+\frac 37+\frac{{\cal
J}^2}{27}+i \frac 12 \sqrt 3 (A^+-A^-)} \cr
\a_3^{\pm} = & \pm M \sqrt{-\frac 12(A^-+A^+)+\frac 37+\frac{{\cal
J}^2}{27}-i \frac 12 \sqrt 3 (A^+-A^-)} \cr} \qquad ,
\eqno(40)
$$
where the following notation has been used
$$
\eqalign{
& A^{\pm} \equiv (a \pm \sqrt b)^{(1/3)} \cr
& a \equiv \frac 8{343}-\frac{106}{441} {\cal
J}^2+\frac{131}{3402} {\cal J}^4+\frac {1}{19683} {\cal J}^6 \cr
& b \equiv -\frac {256}{50421}{\cal
J}^2+\frac{21284}{583443} {\cal J}^4+\frac {71546}{6751269} {\cal
J}^6+\frac{5309}{4960116} {\cal J}^8+\frac{1}{177147} {\cal
J}^{10} \cr} 
\eqno(41)
$$
If $\left |{\cal J}\right |<1 $, which is realistic enough, then
$b>0$ and $a\pm\sqrt b>0$, and so, there are two real roots and
two  pairs of complex conjugated roots. For the sake of concision
 we will dispense with the expression resulting for the Ernst's potential.

The coefficients $m_k$  of this exact solution can be calculated
by using expressions in [11] to obtain the following ten coefficients
$$
\eqalign{
m_0 = & M \coma m_1 = {\cal J} M^2 \coma m_2=m_3 =0 \coma \cr
m_4 = & -\frac 17 M^5 {\cal J}^2 \coma m_5 =  \frac1{21} M^6
{\cal J}^3 \cr
m_6 = & - M^7 (\frac{1}{49} {\cal J}^2+\frac{1}{63} {\cal J}^4)\cr
m_7 = &  M^8 (\frac{5}{147} {\cal J}^3+\frac{1}{189} {\cal
J}^5)\cr
m_8 = & M^9 (-\frac{1}{343} {\cal J}^2-\frac{1}{49} {\cal
J}^4-\frac{1}{567} {\cal J}^6 )\cr
m_9 = &  M^{10} (\frac{3}{343} {\cal J}^3+\frac{13}{1323} {\cal
J}^5+\frac{1}{1701} {\cal J}^7 ) \cr
m_{10} = & -M^{11} ( \frac{1}{2401} {\cal J}^2+\frac{11}{1029}
{\cal J}^4-\frac{17}{3969} {\cal J}^6+\frac{1}{5103} {\cal J}^8
) \cr}
\eqno(42)
$$
According to these expressions the multipole moments are
$$
\matrix{
M_0 = M  \coma &
M_1 = M^2 {\cal J} \coma &
M_2 = 0 \coma &
M_3 =  0 \coma &
M_4 =  0 \cr}
$$
$$
\matrix{
 \ M_5 =  0 \ , \ &
M_6 =  -M^7 (-\frac{4}{147} {\cal J}^2+\frac{1}{63} {\cal J}^4)
\ , \ &
 M_7 =   M^8 (-\frac{12}{539} {\cal J}^3+\frac{1}{189} {\cal
J}^5)
\cr
}$$
$$
\eqalign{
M_8 = & \ M^9 (-\frac{32}{3773} {\cal J}^2+\frac{554}{63063} {\cal
J}^4-\frac{1}{567} {\cal J}^6)
\cr
M_9 = & - M^{10} (-\frac{26912}{2501499} {\cal
J}^3+\frac{9158}{3216213} {\cal J}^5-\frac{1}{1701} {\cal J}^7)
\cr
M_{10} = &- M^{11} (-\frac{13392}{10081799} {\cal
J}^2+\frac{55500}{15842827} {\cal J}^4-\frac{130}{1281987} {\cal
J}^6+\frac{1}{5103} {\cal J}^8)
\cr }
\eqno(43)
$$
We can see that, by construction, this solution obviously posseses a
higher number of null multipole moments than the previous solution.
Besides, the first multipole moment different from zero, i.e.,
$M_6$, turns out to be proportional to ${\cal J}^2$ again, that is to say,
 one order less than the angular momentum. Nevertheless, it must
be pointed out that its magnitude  is not necessarily smaller than the
first moment different from zero ($M_4$) in the previous case. In
fact, first multipole moment different from zero for each case is
always proportional to ${\cal J}^2$, and so, we can achive striking
solutions in this process which, in comparisson with previous solutions in the
sequence, possess moments of a higher multipole order and higher
magnitude at the same time.

In order to illustrate the behaviour of the multipole moments in
this sequence of exact solutions,  we write out the moment of a
certain multipole order for each solution. By way of example, we will
 compare the moment $M_6$ of each solution with just the
same moment for the case $N=1$, i.e., Kerr's metric
$$
\eqalign{
& M_6^{(1)} = M^7 {\cal J}^6 = M_6^{Kerr}\cr
& M_6^{(2)} = -\frac{1}{33} M^7 {\cal J}^2 = -\frac{1}{33} M_6^{Kerr} \frac
{1}{{\cal J}^4}\cr
& M_6^{(3)} = -\frac{1}{49} M^7 {\cal J}^2-\frac{1}{63} M^7
{\cal J}^4 =-\frac{1}{49} M_6^{Kerr} \frac
{1}{{\cal J}^4}-\frac{1}{63} M_6^{Kerr} \frac
{1}{{\cal J}^2}\cr}
\eqno(44)
$$
That is to say, the higher approximation degree in the series
of solutions, the higher magnitude of the multipole moment. According to
 (44), the sequence of solutions should have a good
behaviour, i.e., progressive diminution of the magnitude of any moment,
  if the parameter ${\cal J}$ were larger than $1$ (absolute
value), i.e.,  $J >M^2$, which is not an expected condition for
any realistic object. 

For these reasons, and others which will
be disscussed in next section we will introduce a different
approach to the Monopole-Dynamic Dipole stationary solution.

\vskip 1cm

\noindent {\bf 4. STATIONARY APROXIMATE M--J SOLUTION }

\vskip 0.5cm

According to the previous section, the construction of an exact
stationary and axisimetric solution by Sibgatullin's
method requires the structure of Ernst's potential on the
symmetry axis as a polynomial ratio. Nevertheless, we have shown
that the structure of the Ernst's potential on the symmetry axis
corresponding to a solution of type $M-J$ is, in some way,
 a ratio of series, which means that it
cannot be expresed as  a polynomials ratio. Hence, although the
potential of such a solution were obtained on the symmetry axis,
we cannot apply  Sibgatullin's method. Besides, the sequence
of exact solutions previously proposed approches the $M-J$
solution in a rather unexpected way, since the magnitude of
multipole moments does not decrease while the approximation degree
raises. 

Therefore, in this section we will proceed 
 to approach the $M-J$ solution in a different manner. We give
up looking for exact solutions and propose instead a sequence of approximate
solutions as partial sums of  power series on the parameter ${\cal
J}$.

The expressions in (3) of the coefficients $m_n$ corresponding to the
solution $M-J$ lead to an  Ernst's potential $\xi$ on the symmetry
axis as a power series in the parameter ${\cal J}$.
According to this result we will look for solutions
in that way. Let us consider the Ernst's equation for the 
potential $\xi$ and let us assume a solution  of the form
$$
\xi \equiv \xi_0+\sum_{\a=1}^{\infty} \xi_{\a} {\cal J}^{\a}
\coma
\eqno(45)
$$
where $\xi_0$ represents the the Ernst's potential corresponding to 
Schwarzschild's solution. Imposing this series to verify
Ernst equation at each order leads to the following equations
concerning the functions $\xi_{\a}$
$$
\eqalign{
& (\xi_0^2-1) \triangle\xi_{2\a+1}-4 \xi_0 \nabla\xi_0
\nabla\xi_{2\a+1}+2 \xi_{2\a+1} (\nabla \xi_0)^2 = H_{2\a+1} \cr
\vspace{2mm}
& (\xi_0^2-1) \triangle\xi_{2\a}-4 \xi_0 \nabla\xi_0
\nabla\xi_{2\a}+2 \xi_{2\a} (\nabla \xi_0)^2
\frac{\xi_0^2+1}{\xi_0^2-1} = H_{2\a}
\cr}\coma
\eqno(46)
$$
where first equation refers to odd orders and second one to
even orders ($\a = 1,2,\dots$); the second members of those equations
are given by
$$
 H_{\a} = \sum_{\scriptstyle i+j+k=\a \atop\ i,j,k < \a } (-1)^i
\[2 \xi_i
\nabla\xi_j
\nabla
\xi_k-\xi_i
\xi_j \triangle \xi_k \] \quad , \quad \a >0 \quad .
\eqno(47)
$$
i.e., equation of order $\a$ depends on the previous orders.

The previous equations can be simplified  by redefining the functions
$\xi_{\a}$ as follows
$$
\z_{\a} \equiv \frac{\xi_{\a}}{\xi_0^2-1} \quad .
\eqno(48)
$$
which leads at each order $\a$ to the following equations
$$
(\xi_0^2-1)\triangle \z_{\a}+\xi_0^2 \nu^2 \z_{\a} =
\frac {H_{\a}} {\xi_0^2-1} \coma
\left\{\eqalign{
\nu & =0 \coma \a = {\rm even} \cr
\vspace{2mm}
\nu & =2 \coma \a = {\rm odd} \cr
}\right.
\eqno(49)
$$
It is easy to solve this equation by writing it in prolate
coordinates. As usual the general solution can be obtained by
adding a particular solution of inhomogeneous equation to the
general solution of whole equation. Moreover we impose a regular
behaviour  on the symmetry axis ($y=\pm 1$)  at least like
$1/x$ with respect to the variable
$z$ in the neighbourhood of infinity. We then obtain:
$$
\eqalign{
&\xi=\frac 1x+\sum_{\a =1}^{\infty} {\cal J}^{\a} \xi_{\a}(x,y) \cr
\vspace{2mm}
&\xi_{\a}(x,y)=\frac{x^2-1}{x^2} \[\z_{\a}^{P}(x,y)+ \sum_{n =
0}^{\infty} h_n^{\a} Q_{n}^{(\nu)}(x) P_n(y) \] \cr
}
\quad ,
\eqno(50)
$$
where $Q_{n}^{(\nu)}(x)$ are associated Legendre's functions of
second kind, the functions $\z_{\a}^{P}(x,y)$ are particular solutions
of the inhomogeneous equations corresponding to each order $\a$,
and $h_n^{\a}$ are arbitrary constants.

To describe the Monopole-Dynamic dipole solution ($M-J$) from
the general solution in (50), it is necessary to add as a boundary
condition the behaviour of potential $\xi$ on the symmetry
axis, which has been defined previously by the series in (5). Hence, we
force now the function $\Phi_{\a}$ appearing in (5) to agree with
the corresponding restriction on the symmetry axis of the function
$\xi_{\a}$ of general solution (50), which  leads to determine the 
constants $h_n^{\a}$.

Previously, the functions $\Phi_{\a}$  must be adapted to the
structure of the general solution in (50), and so, we begin by taking
 a factor $\displaystyle
\frac{M^2-z^2}{z^2}$
out of the expression of the function $\Phi_{\a}$,
$$
\Phi_{2n} = \frac{M^2-z^2}{z^2} \[ \frac{1}{\hat \lambda^2-1}
\sum_{k=2n}^{\infty} G(2n,2k) \hat\lambda^{2k+1} \] \quad .
\eqno(51)
$$
By carrying out an expansion on the
parameter
$\hat\lambda$ results in
$$
\Phi_{2n} = -\frac{M^2-z^2}{z^2} \sum_{i=0}^{\infty}
\sum_{k=2n}^{\infty} G(2n,2k) \hat\lambda^{2k+2i+1} \coma
\eqno(52)
$$
that is to say,
$$
\Phi_{2n} = -\frac{M^2-z^2}{z^2}
\sum_{j=2n}^{\infty}\hat\lambda^{2j+1}
\sum_{k=2n}^{j} G(2n,2k) \quad .
\eqno(53)
$$

Below we write these functions $\Phi_{2n}$ in terms of the
 Legendre's functions of second kind, by using Lemma 4 of the
Appendix
$$
\Phi_{2n} = -\frac{M^2-z^2}{z^2} \sum_{j=2n}^{\infty}
I_j^{(2n)}
\sum_{i=j}^{\infty} (4i+1) L_{2i,2j} Q_{2i}(1/\hat\lambda) \coma
\eqno(54)
$$
where the following notation has been used 
$$
I_j^{(2n)} \equiv \sum_{k=2n}^{j} G(2n,2k) \quad .
\eqno(55)
$$

With respect to the odd functions $\Phi_{2n+1}$ we proceed in the
same way to obtain mentioned factor and write, in this case, these 
functions in terms of  associated Legendre's functions of second
kind $Q_{2l+1}^{(2)}(1/\hat\lambda)$, which can be obtained making
use of Lemma 5 presented in the Appendix, and so,
$$
\Phi_{2\a+1} = \frac{z^2-M^2}{z^2} \sum_{l=2\a}^{\infty}
\frac{4l+3}{(2l+2)(2l+1)} Q_{2l+1}^{(2)}(1/\hat
\lambda)
\sum_{n=2\a}^{l} \frac{2l+2n+1}{(2n+1)!!}
L_{2l,2n} I_n^{(2\a+1)} 
\eqno(56)
$$
with the following notation
$$
I_n^{(2\a+1)} \equiv \sum_{k=2\a}^n G(2\a+1,2k+1) \quad .
\eqno(57)
$$

At this point we proceed to determine the constants $h_n^{\a}$ of
the general solution (50) which  correspond to the $M-J$
solution . In that way, we choose particular solutions of
inhomogeneous equations (49) as follows
$$
\eqalign{
\z_{2\a}^{P} = & \sum_{l=0}^{\infty} (4l+1) Q_{2l}(x) S_{2l}(y)
\cr
\vspace{2mm}
\z_{2\a+1}^{P} = & \sum_{l=0}^{\infty} (4l+3) Q_{2l+1}^{(2)}(x)
S_{2l+1}(y) \cr} \coma
\eqno(58)
$$
 $S_{a}(y)$ being polynomials  in  the angular variable.

Comparing general solution (50), evaluated on the
symmetry axis, with the previous expressions (54) and (56) gives
$$
 \eqalign{
& h_{2n+1}^{2\a} =   0 \cr
\vspace{2mm}
& h_{2n}^{2\a} = -(4n+1) S_{2n}(1) \coma n< 2\a \cr
\vspace{2mm}
& h_{2n}^{2\a} = -(4n+1) \[ S_{2n}(1)+\sum_{k=2\a}^n L_{2n,2k}
 I_k^{(2\a)} \] \coma n \geq 2\a \cr
\vspace{2mm}
& h_{2n}^{2\a+1} = 0 \cr
\vspace{2mm}
& h_{2n+1}^{2\a+1} = -(4n+3) S_{2n+1}(1) \quad, \quad n< 2\a \cr
\vspace{2mm}
& h_{2n+1}^{2\a+1} = -(4n+3) \[S_{2n+1}(1)+
\frac{1}{(2n+1)(2n+2)} \sum_{l=2\a}^n
 \frac{2n+2l+1}{(2l+1)!!} L_{2n,2l}
I_l^{(2\a+1)} \] \cr
& \hskip 11cm n \geq 2\a \cr}
\eqno(59)
$$

Let us construct explicitely  the first orders of the
solution $M-J$. Obviously, the order zero contribution to the solution
must be Schwarzschild's solution, since taking the parameter
${\cal J}=0$ leads us to consider the mass as the unique multipole
moment. The Ernst's potential $\xi$ of Schwarzschild's solution
$(\xi_0=1/x)$ equals the structure described on the symmetry axis
(5).
\vskip 3mm

{\bf A) FIRST ORDER.}
\vskip 2mm
The first contribution on the parameter ${\cal J}$ should be a
solution of the first equation in (46) turns out to be
homogeneous at order one, and hence
$$
\xi_1 = \frac{1-x^2}{x^2} \sum_{l=0}^{\infty} h_l^1 Q_l^{(2)}(x)
P_l(y) \quad .
\eqno(60)
$$
This expression on the symmetry axis, gives
$$
\xi_1(y=1) = 2 \hat\lambda h_0^1+2 \hat\lambda^2
h_1^1+(\hat\lambda^2-1)^2
\sum_{l=2}^{\infty}h_l^1 Q_l^{(2)}(1/\hat\lambda) \quad .
\eqno(61)
$$
In addition, the first contribution to the $M-J$ solution  must 
agree with (5) on the symmetry axis and so, the only solution
corresponds to the following choice of constants
$$
\matrix{
h_0^1 =  0  \coma &
h_1^1 =  \frac12 \coma &
h_l^1 =  0 , \quad l \geq 2 \cr} \coma
\eqno(62)
$$
which means,
$$
\xi_1(x,y) = \frac12 \frac{1-x^2}{x^2} Q_1^{(2)}(x) P_1(y) =
\frac{y}{x^2} \quad .
\eqno(63)
$$
It must be noted that this approximate solution is the same as the
one arising from the expansion of Kerr's metric on
parameter ${\cal J}$ up to the first order.
\vskip 3mm
{\bf B) SECOND ORDER.}
\vskip 2mm
A particular solution of the inhomogeneous equation (46),
corresponding to order two, results in
$$
\z_2^{inh} = (\frac{x}{2}-\frac{y^2}{x}) \frac{1}{1-x^2} \quad .
\eqno(64)
$$
and in view of Lemma 4 of the Appendix, this could be
rewritten in terms of associated Legendre's functions in the following
way
$$
\z_2^{inh} = \sum_{n=0}^{\infty} Q_{2n}(x) (4n+1)
\[\frac 12-y^2(1-L_{2n,0})\] \quad .
\eqno(65)
$$
By substituing this particular solution in the expressions
(59) we determine the constants of the contribution of the second
 order to the solution, i.e., 
$$
\eqalign{
h_0^2 = & \frac12  \coma h_2^2 = 5\(L_{2,0}-\frac 12\) \cr
h_{2n}^2 = & (4n+1) \[ -\frac 12+L_{2n,0}-\sum_{k=2}^n L_{2n,2k}
I_k^{(2)}\] \coma  n \geq 2 \cr} \quad .
\eqno(66)
$$
Now, taking the expression $G(2,2j)$ (8) into account  we can
write term  $I_k^{(2)}$ as a sum of irreducible fractions in the
following form
$$
I_k^{(2)} = \frac{(k+3)(k-1)}{(2k+3)(2k+1)} =
\frac 14-\frac{15}{8} \frac{1}{2k+1}+\frac{15}{8} \frac{1}{2k+3} \quad .
\eqno(67)
$$
which implies that  $I_0^{(2)}=-1$ and $I_1^{(2)}=0$. Therefore
constants $h_{2n}^2$ with $n \geq 2$ can be written as follows
$$
h_{2n}^2 = (4n+1) \[ -\frac 12-\sum_{k=0}^n L_{2n,2k}
I_k^{(2)}\] \quad .
\eqno(68)
$$

Considering now expression (67) we have
$$
h_{2n}^2 = (4n+1) \[
-\frac 12+\frac{15}{8}\sum_{k=0}^n\frac{L_{2n,2k}}{2k+1}-
\frac{15}{8}\sum_{k=0}^n\frac{L_{2n,2k}}{2k+3}-\frac
14 \sum_{k=0}^n L_{2n,2k}\] \ .
\eqno(69)
$$
and making use of Lemma 2 of the Appendix, and  the
ortonormality of the Legendre's polynomials, we have finally the
following expressions
$$
\matrix{
h_0^2 =  \frac 12 \coma &
h_2^2 =  -5  \coma &
h_{2n}^2 =  -\frac 34 (4n+1) , \quad n \geq 2 \cr} \quad .
\eqno(70) 
$$
Hence, the contribution of order $2$ to the solution $M-J$ can be
written as follows
$$
\eqalign{
\xi_2(x,y) = \frac 1{2x}-\frac{y^2}{x^3}+ &\frac{1-x^2}{x^2}
[\frac 1 2 Q_0(x)P_0(y)-5 Q_2(x) P_2(y)- \cr
\vspace{2mm}
& -\sum_{n=2}^{\infty}
\frac 3 4  (4n+1) Q_{2n}(x) P_{2n}(y) ] \quad ,\cr}
\eqno(71)
$$
an expression which can be simplfied by making use of Heine's
identity [14]:
$$
\frac{x}{x^2-y^2} = \sum_{n=0}^{\infty}(4n+1) Q_{2n}(x) P_{2n}(y) \coma
\eqno(72)
$$
which leads to the following expression
$$
\xi_2(x,y) = \frac 1{2x}-\frac{y^2}{x^3}+\frac{1-x^2}{x^2}
\[\frac 54 Q_0(x)P_0(y)-\frac54 Q_2(x) P_2(y)-\frac 34
\frac{x}{x^2-y^2}
\] \ ,
\eqno(73)
$$
that is to say
$$
\xi_2(x,y) = \frac 54 \frac{x^2-1}{x^2}
\[ Q_0(x)P_0(y)- Q_2(x) P_2(y)-
\frac{x}{x^2-y^2}
\]+\xi_2^{Kerr} \ ,
\eqno(74)
$$
where $\xi_2^{Kerr}$ is the order $2$ in the expansion of Kerr's metric for
the parameter ${\cal J}$.

The multipole moments can be calculated by
$FHP$ algorithm from coefficients $m_n$. For the  solution up to
order $1$, i.e., $\xi_{M-J}^{(1)}\equiv\xi_0+{\cal J} \ \xi_1$, all
coefficients $m_n$ with
$n\geq 2$ are zero (i.e., of higher order than ${\cal J}$). So,
its multipole moments equal those in the exact
solution presented in the previous section whose potential $e(z)$
was a ratio of polynomials of order $2$, that is to say,
$$
\eqalign{
M_0 =& M \coma 
M_1 =  {\cal J} M^2 \coma 
M_2 = 0 \coma 
M_3 = 0 \coma \cr 
\vspace{2mm}   
M_4 =&  \frac 17 {\cal J}^2 M^5 \coma 
M_5 = -\frac 3{21} {\cal J}^3 M^6 \coma 
M_6 = -\frac 1{33} {\cal J}^2 M^7 \cr
\vspace{2mm}
M_7 =&  \frac {19}{429} {\cal J}^3 M^8 \coma 
M_8 =  - M^9 \(-\frac 1{143} {\cal J}^2
+\frac{53}{3003} {\cal J}^4 \) \cr
\vspace{2mm}
M_9 =&   M^{10} \( -\frac{43}{2431}
{\cal J}^3 + \frac{41}{17017} {\cal J}^5 \) \coma 
M_{10} =   M^{11} \(-\frac 7{4199} {\cal
J}^2+\frac{202}{12597} {\cal J}^4 \).\cr} 
\eqno(75)
$$
With respect to the $M-J$ solution up to the second order,
i.e., $\xi_{M-J}^{(2)} \equiv \xi_0+{\cal J} \ \xi_1+{\cal J}^2 \
\xi_2 $, we have the following multipole moments
$$
\eqalign{
M_0 & = M \coma
M_1  =  {\cal J} M^2 \coma
M_2  = 0 \coma
M_3  = 0 \coma     
M_4  = 0 \cr
\vspace{2mm}
M_5 & = - \frac 1{21} {\cal J}^3 M^6 \coma
M_6  = 0 \coma
M_7  =  - \frac {59}{3003} {\cal J}^3 M^8 \coma
M_8  =  - \frac 2{231} {\cal J}^4 M^9 \cr
\vspace{2mm}
M_9 & =   M^{10} \( \frac{41}{17017}
{\cal J}^5 + \frac{593}{51051} {\cal J}^3 \) \coma
M_{10}  =   - \frac {49873}{6789783} {\cal
J}^4 M^{11}  \cr} 
\eqno(76)
$$

It can be seen from (75) and (76) that the higher the order of
approximation to  $M-J$ solution, the higher (one order more) the
order in the parameter ${\cal J}$ of its  non vanishing
multipole moments.

The structure of coefficients $m_k$ in terms of the parameter ${\cal
J}$  shows which is the order $n$ of such coefficients that possess a
contribution of order $\a$ in ${\cal J}$.  In fact,  the first
contribution to an even power in the parameter, arises from the 
coefficient of order $2\a$. If  $\a$ is odd, such
contribution arise from the coefficient of order $2\a-1$. Since  the
relation between the multipole moment and the coefficient $m_k$ of same
order is linear we
can conclude that the multipole
moments of the solution which approche the $M-J$ solution 
up to the order $\a$ in the
parameter ${\cal J}$ have the following
characteristics:
\bigskip

{\bf 1)} If the order $\a$ is even, all its massive multipole moments
up to $M_{2\a+2}$ (incl.) will be zero, the following ones 
being at least
of order $\a+2$ in parameter ${\cal J}$.
With regard to its dynamic multipole moments they will be zero up
to  $M_{2\a-1}$ (incl.), and the following ones will be at least of
order $\a+1$ in parameter ${\cal J}$.

\medskip

{\bf 2)} If the order of approximation $\a$  is odd, then all
massive multipole moments  will be zero up to  $M_{2\a}$
(incl.) and the ones higher than that will be at least of order $\a+1$, 
whereas the
dynamic moments will be zero up to $M_{2\a+1}$  and the following 
moments will be at least of order $\a+2$.

These results show, in the same way as the static $M-Q$ solution 
 [5], that it is possible to understand the series in (45) in terms of the 
perturbations theory. Each partial sum of that series is a
better approximation to the solution which only has mass and
angular momentum, since the multipole moments higher than this are
 either
zero or have an order in the parameter ${\cal J}$  higher than the
one of the approximation. In addition, unlike the solutions in the 
previous section, the multipole moment of a certain order is
progressively smaller  as the order of approximation in the 
series (45) grows.

\vskip 1cm

\noindent {\bf APPENDIX}
\vskip 2mm
In this Appendix we will enunciate a sequence of Lemmas about
several properties of the Legendre's polynomials and the associated Legendre's
functions of second kind. Results of these Lemmas are probably
well known, but proofs for them can be easily obtained 
considering some results of various Lemmas appearing in [5].

Let us $P_{2n}$ be a Legendre's polynomial of even order in an arbitrary
variable
$$
P_{2n}(\zeta) =\sum_{k=0}^n L_{2n,2k} \zeta^{2k} \coma
\eqno(A.1)
$$
where its coefficients have the following expression
$$
L_{2n,2k}\equiv
(-1)^{n-k}2^{k-n}\frac{(2n+2k-1)!!}{(n-k)!(2k)!} \quad .
\eqno(A.2)
$$
Let us consider the development of  
an arbitrary variable to an even power in terms of Legendre's polynomials
in that variable, i.e.,
$$
\zeta^{2n} = \sum_{k=0}^{\infty} C_{2n,2k} P_{2k}(\zeta) \quad ,
\eqno(A.3)$$
where coefficients $C_{2n,2k}$  can be obtained by integration
from the following expression,
$$
C_{2n,2k}=\frac{4k+1}{2} \int_{-1}^1P_{2k}(\zeta)\zeta^{2n}d\zeta
\quad ,
\eqno(A.4)$$
and so, for several values of  indexes $k$ and 
$n$ such a coefficients turns out to be
$$
C_{2n,2k} =
\left\{\eqalign{
 (4k+1)\frac{2n!}
{(2n-2k)!!(2n+2k+1)!!} \  :\  & k \leq n \cr
 0 \qquad \qquad\  :\  & k > n  \cr
}\right. \quad .
\eqno(A.5)
$$

\vskip 1cm

\noindent \underbar{\bf Lemma 1.\/} \  The following orthogonality
relation is satisfied:
$$
\sum_{j=0}^k L_{2k,2j}C_{2j,2n} = \delta_{kn} \quad .
\eqno(A.6)$$

\vskip 0.5cm

\noindent \underbar{\it Corolary:}  \  It can be deduced
evidently another orthogonality relation
$$
\sum_{k=n}^j L_{2k,2n} C_{2j,2k} = \delta_{jn} 
\eqno(A.7)
$$

\vskip 1cm

\noindent \underbar{\bf Lemma 2.\/}  \  For all pair of  positive
entire numbers $n$ and $k$ such that $n<k$, the following equality is
 verified
$$
\sum_{j=0}^k \frac{L_{2k,2j}}{2n+2j+1} = 0 \quad .
\eqno(A.8)$$

\vskip 1cm

\noindent \underbar{\bf Lemma 3.\/}  \  For all pair of  positive
entire numbers  $\a$ and  $j$ the following equality is
 verified:
$$
\sum_{n=2\a}^{\infty} \frac{\hat \lambda^{2n+1}}{2n+2j+1} = \hat
\lambda^{4\a} \sum_{n=0}^{j+2\a} C_{2j+4\a,2n} Q_{2n}(1/\hat
\lambda) \coma \hat \lambda \equiv \frac Mz \quad .
\eqno(A.9)
$$

\vskip 1cm

\noindent \underbar{\bf Lemma 4.\/}  \  For all positive entire
number $n$ following is verified
$$
\eqalignno{
\frac{1}{x^{2n+1}} = & \sum_{k=n}^{\infty}(4k+1) L_{2k,2n}
Q_{2k}(x) & (A.10a) \cr
\frac{1}{x^{2n}} = & \sum_{k=n-1}^{\infty}(4k+3) L_{2k+1,2n-1}
Q_{2k+1}(x) \quad .& (A.10b) \cr}
$$

\vskip 1cm

\noindent \underbar{\bf Lemma 5.\/}  \  For all positive entire
number $n$ following is verified
$$
\frac{1}{x^{2n}} = \frac1{(2n-1)!!} \sum_{k=n-1}^{\infty}
(4 k+3) Q_{2k+1}^{(2)}(x)  L_{2k,2n-2}
\frac{2k+2n-1}{(2k+2)(2k+1)} \quad .
\eqno(A.11)
$$

\vskip 1cm

\noindent {\bf REFERENCES}

\noindent [1] Fodor, G., Hoenselaers, C., Perj\'es, Z., 1989 \quad {\it J. Math.
Phys.} {\bf 30}, 2252

\noindent [2] Sibgatullin, N.R., 1984 \quad {\it Oscillations and
waves in strong gravitational and electromagnetic fields},
Nakua, Moscow. (English traslation: Springer-Verlag Ed. Berlin,
1991)

\noindent [3] Geroch, R. 1970 \quad {\it  J. Math. Phys.} {\bf 11},
2580
 
\noindent [4]  Hansen, R.O., 1974 \quad {\it J. Math. Phys.} {\bf
15}, 46

\noindent[5] Hern\'andez-Pastora, J.L., Mart\'\i n, J., 1994
\quad {\it Gen. Rel. Grav.} {\bf 26}, 877.

\noindent [6] Manko, V.S., Sibgatullin, N., 1993  \quad {\it
Class. Quantum Grav.} {\bf 10}, 1383

\noindent \quad \ Manko, V.S., Mart\'\i n, J., Ruiz, E., 1993  \quad
{\it Phys. Lett. A} {\bf 196}, 23

\noindent [7] Herrera, L., Manko, V.S., 1992  \quad
{\it Phys. Lett. A} {\bf 167}, 238

\noindent [8] Manko, V.S., Mart\'\i n, J., Ruiz, E., Sibgatullin, N.R., Zaripov,
M.N., 1994 \quad {\it Phys. Rev. D} {\bf 49}, 5144

\noindent [9]  Ruiz, E. Manko, V.S., Mart\'\i n, J., 1995 \quad {\it
Phys. Rev. D} {\bf 51}, 4192

\noindent [10] Ernst, F.J. 1968\quad {\it   Phys. Rev.} {\bf 167},
1175.

 \noindent  \qquad Ernst, F.J. 1968 \quad {\it Phys. Rev.} {\bf
168}, 1415

\noindent [11] Manko, V.S. and Ruiz, E., 1998 \quad {\it Class. Quantum Grav.}
(to appear)

\noindent [12] Kordas, P., 1995 \quad {\it Class. Quantum Grav.}
{\bf 12}, 2037

\noindent [13] Meinel, R., Neugebauer, G., 1995 \quad {\it Class.
Quantum Grav.} {\bf 12}, 2045

\noindent [14] Lense, J., 1947 \quad {\it Reihenentwicklung in der 
Mathematischen Physik}, Walter de Gruyter and Co.. Berlin.

 \bye